\documentclass[fleqn,12pt]{wlscirep}
\usepackage{amsmath}
\usepackage{graphicx}
\usepackage{footmisc}
\usepackage{color}

\begin{document}

\title {Gapped electron liquid state in the symmetric Anderson lattice, Kondo insulator state}
\author[1]{Igor N. Karnaukhov}
\affil[1]{G.V. Kurdyumov Institute for Metal Physics, 36 Vernadsky Boulevard, 03142 Kiev, Ukraine}
\affil[*]{karnaui@yahoo.com}
\begin{abstract}
The Kondo insulator state (KIS) realized in the symmetric Anderson model at half filling is studied in the framework of a mean field approach.
It is shown that the state of the Kondo insulator is realized in a lattice with a double cell and a gapped electron liquid behaves like a gapless
Majorana spin liquid. The local moments of d-electrons form a static $Z_2$-field in which band electrons move. The gap value in the quasi-particle excitations spectrum decreases with increasing an external magnetic field and closes at its critical value. The behavior of an electron liquid is studied for an arbitrary dimension of the model. The proposed approach leads to the description of KIS without the need to resort to artificial symmetry breaking to alternative understanding of the physical nature of this phase state.
\end{abstract}
\maketitle

\section*{Introduction}

As a rule  in the traditional interpretations of the Kondo lattice and Anderson lattice models  interaction is decoupled in favour of a hybridization of electrons with different spins \cite{CA,IK1}. However, such an approximation breaks the local gauge symmetry: it does not conserve the spins of s- and d-electrons separately in the Anderson lattice, does not conserve the total number of band electrons in the Kondo lattice since the local d-occupation is no longer conserved.
The effective Hamiltonian is obtained within a mean field approach, determines  the ground state and low-energy quasi-particle excitations
of the electron liquid.  The effective Hamiltonian must have  the same symmetry as the Hamiltonian of the model,
otherwise the  mean field approach can not be considered as adequate.

The Hamiltonian of the Anderson model has an exact solution when there is no hybridization between electrons or the Hubbard repulsion.
Therefore, it is necessary  to take into account both hybridization and repulsion between electrons for problem solving.
This makes it possible to explicitly take into account the scattering of band electrons by local moments with spin flip.
The Kondo effect in the Kondo problem and KIS in the Kondo lattice are realized  to this scattering.
In  \cite{e1} a compound $FeSB_2$ as a candidate topological Kondo insulator based on $3d$-electrons is investigated. The authors believe
that insulator state similar to those in $SmB_6$ \cite{e2}, $YbB_{12}$ \cite{e3}  exist in the compound with $d-$ instead of $f-$electrons also.
The Anderson lattice model describes non topological KIS, so a traditional KIS will be studied.

The purpose of the paper is  KIS  realized in the symmetric Anderson lattice
within the mean field approach without  local gauge symmetry breaking of the model studying .
A gapped electron liquid behaves like  a gapless Majorana spin liquid in the Kitaev model \cite{AK,IK2,IK3} in KIS.
The local moments of d-electrons form a static $Z_2$-field in which band electrons move.
A configuration of the $Z_2$-field corresponding to the ground state forms the lattice
with a double cell and a global gauge symmetry is not broken.

\section*{Model}

The Hamiltonian of the Anderson lattice is the sum of two terms, the first one determines the energies of  s- and d-electrons and hybridization between them, the second one takes into account the on-site repulsion of d-electrons ${\cal H}={\cal H}_{0}+{\cal H}_{int}$
\begin{eqnarray}
&&{\cal H}_0= -
\sum_{<i,j>}\sum_{\sigma=\uparrow,\downarrow}
c^\dagger_{i,\sigma} c_{j,\sigma} +
v\sum_{j=1}^{N}\sum_{\sigma=\uparrow,\downarrow}
(c^\dagger_{j,\sigma} d_{j,\sigma}+ d^\dagger_{j,\sigma} c_{j,\sigma})+\nonumber\\&&
(\epsilon_g+\frac{1}{2}U) \sum_{j=1}^{N}\sum_{\sigma=\uparrow,\downarrow}n_{j,\sigma}-
H \sum_j (n_{j, \uparrow}-n_{j,\downarrow}+c^\dagger_{j,\uparrow} c_{j,\uparrow}-c^\dagger_{j,\downarrow} c_{j,\downarrow}),
         \nonumber\\&&
{\cal H}_{int}= U\sum_{j=1}^{N}\left( n_{j,\uparrow}-\frac{1}{2} \right) \left( n_{j,\downarrow}-\frac{1}{2} \right),
\label{eq:H}
\end{eqnarray}
where $c^\dagger_{j,\sigma},c_{j,\sigma}$ and $d^\dagger_{j,\sigma},d_{j,\sigma}$ ($\sigma=\uparrow,\downarrow)$ are the fermion operators determined at a lattice site $j$, $U$ is the  value of the on-site Hubbard interaction determined by the density operator $n_{j,\sigma}=d^\dagger_{j,\sigma}d_{j,\sigma}$, the band width of s-electrons is determined by the hopping integral equal to unity, the energy of flat band of d-electrons is equal to $\epsilon_g$,  $v$ determines the hybridization of s- and d-electrons, $H$ is an external magnetic field, N is the total number of atoms.

   \begin{figure}[tp]
        \centering{\leavevmode}
   \begin{minipage}[h]{.45\linewidth}
   \center{
   \includegraphics[width=\linewidth]{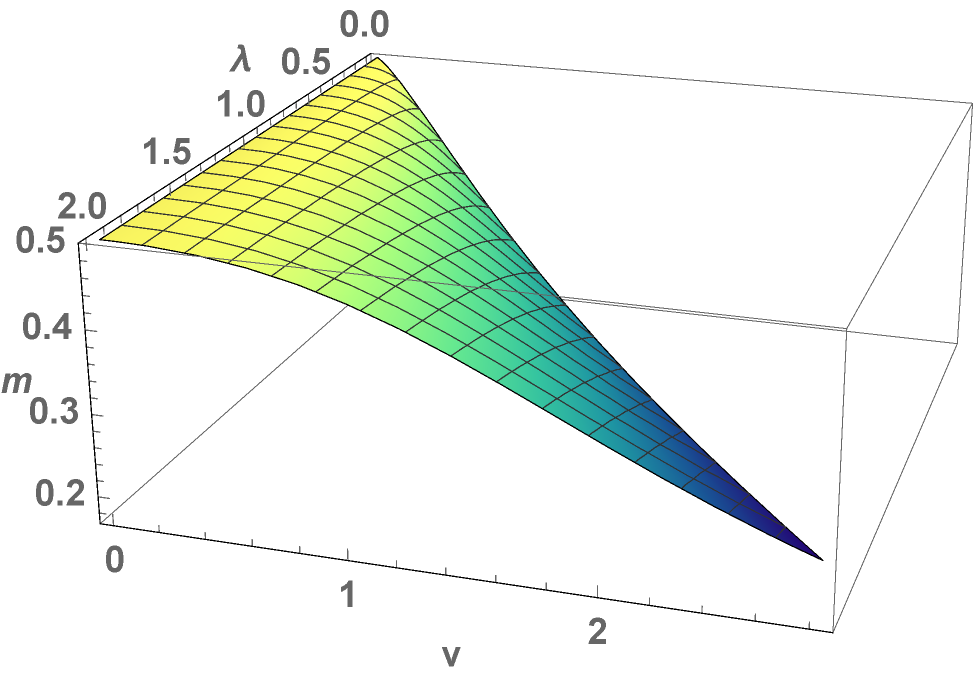}\\
   }
    \end{minipage}
\caption{(Color online) A local moment of d-electrons $m=\frac{1}{2N}\sum_{\textbf{k}}(n_{\textbf{k},\uparrow}-n_{\textbf{k},\downarrow})$
     as  a function of $\lambda, v$, calculated for $v<2\sqrt{2} \lambda$ for a square lattice at half filled occupation,
          a value of a local moment changes from $\frac{1}{2}$ at $v=0$ to $\frac{1}{6}$ at $v\to\infty$.
              }
\label{fig:0}
     \end{figure}

    \begin{figure}[tp]
         \centering{\leavevmode}
     \begin{minipage}[h]{.45\linewidth}
     \center{
     \includegraphics[width=\linewidth]{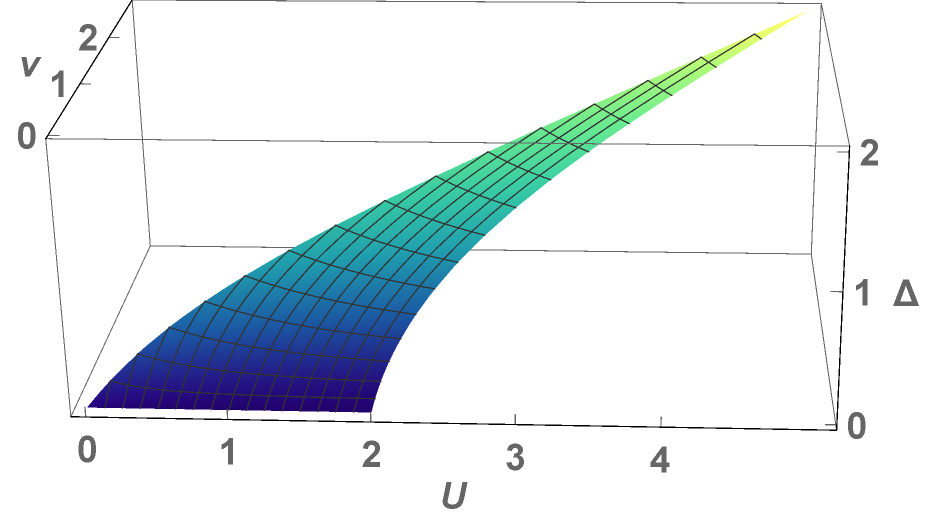} a)\\
                      }
        \end{minipage}
     \begin{minipage}[h]{.45\linewidth}
     \center{
     \includegraphics[width=\linewidth]{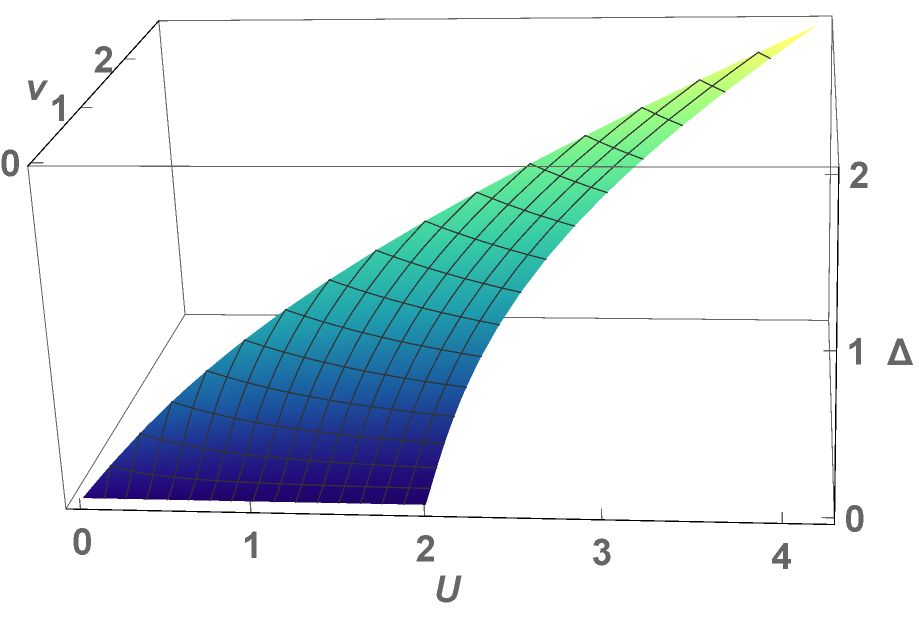} b)\\
                      }
        \end{minipage}
        \begin{minipage}[h]{.45\linewidth}
     \center{
     \includegraphics[width=\linewidth]{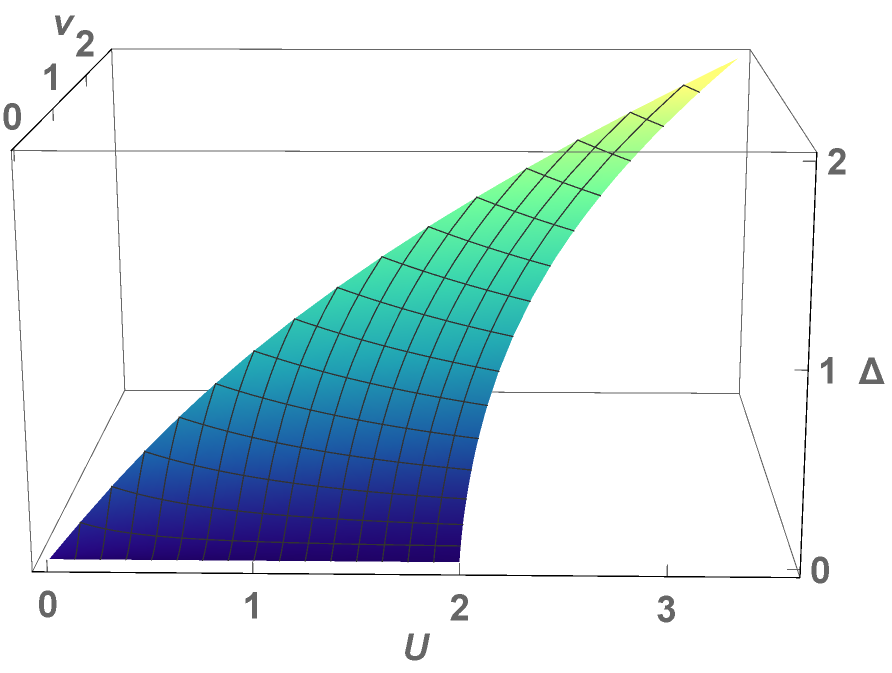} c)\\
                      }
        \end{minipage}
     \caption{(Color online)
     A value of the gap $\Delta (0)$ as a function of $U$ and $v$  calculated for the chain a), square b)  and cubic c) lattices under the condition
     $v<2\sqrt{ 2}\lambda$.
       }
     \label{fig:3}
     \end{figure}

\begin{figure}[tp]
  \centering{\leavevmode}
\begin{minipage}[h]{.45\linewidth}
\center{
\includegraphics[width=\linewidth]{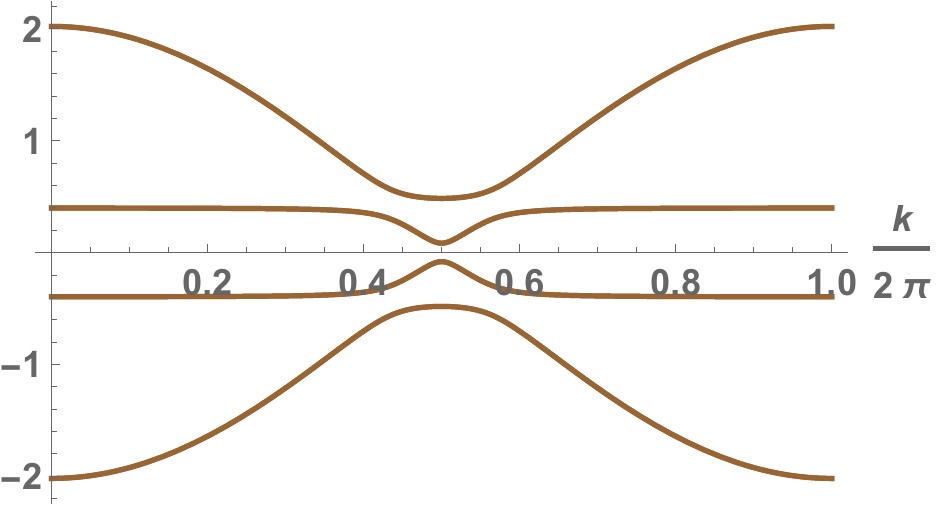} a)\\
                 }
   \end{minipage}
\begin{minipage}[h]{.45\linewidth}
\center{
\includegraphics[width=\linewidth]{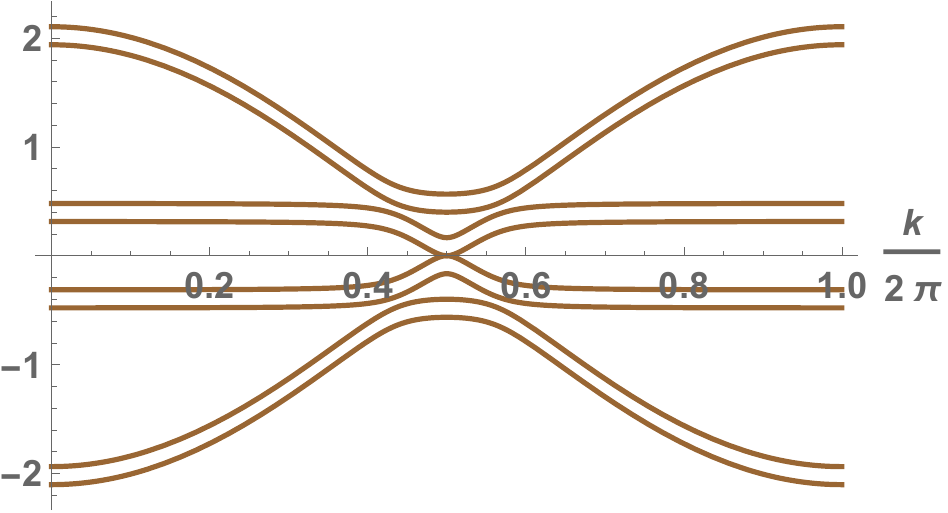} b)\\
                 }
   \end{minipage}
\caption{(Color online)
 A spectrum  of the symmetric Anderson chain as  a function of the wave vector calculated under the condition $v<2\sqrt{2} \lambda$ for $v=0.2$,  $\lambda=0.2$:  $\Delta=0.1656$ at $H=0$  a), $\Delta= 0$ at $H_c=0.0828$  b).
  }
\label{fig:1}
\end{figure}

\begin{figure}[tp]
     \centering{\leavevmode}
\begin{minipage}[h]{.45\linewidth}
\center{
\includegraphics[width=\linewidth]{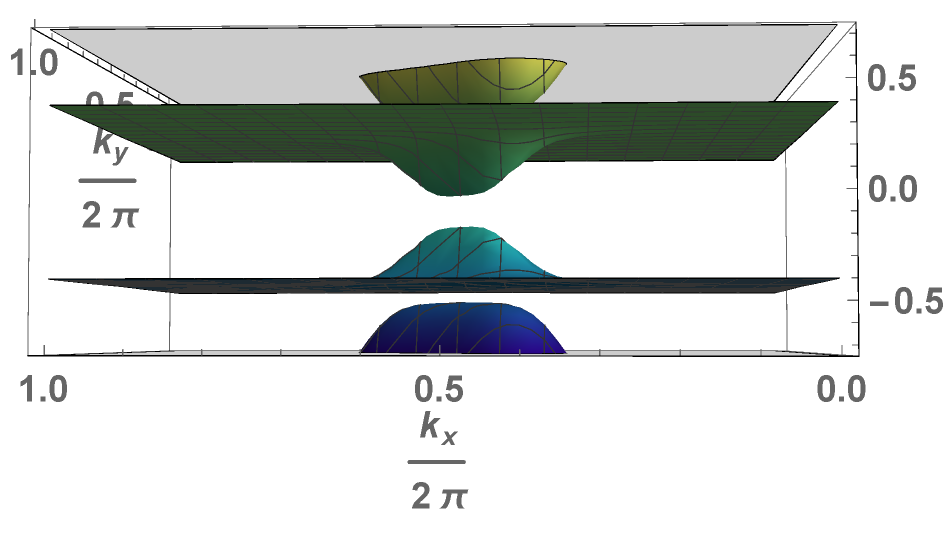} a)\\
                 }
   \end{minipage}
\begin{minipage}[h]{.45\linewidth}
\center{
\includegraphics[width=\linewidth]{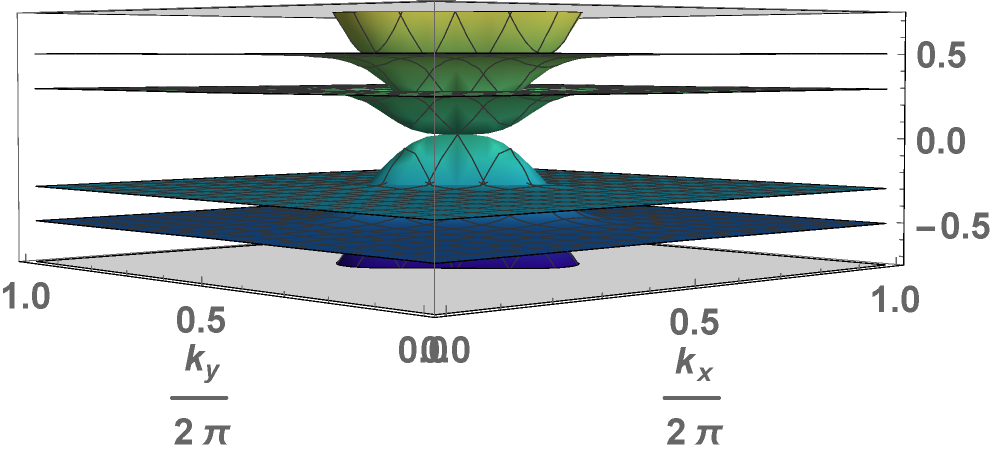} b)\\
                 }
   \end{minipage}
   \begin{minipage}[h]{.32\linewidth}
\center{
                 }
   \end{minipage}
\caption{(Color online)
 A  low energy part of the spectrum  of the symmetric Anderson square lattice  as  a function of the wave vector calculated under the condition
  $v<2\sqrt{2} \lambda$ for $v=0.2$,  $\lambda=0.2$: $\Delta=0.1656$ at $H=0$  a),
$\Delta= 0$ at $H_c=0.0828$ b).
  }
\label{fig:1}
\end{figure}

The electron liquid behavior in the Kondo problem and the Anderson model is universal
 in the sense that it does not depend on the dimension of the system
and their exact solutions do not depend on dimension  of the models \cite{W1,W2}.
We will show that this universality is preserved for the symmetric Anderson lattice model, as it was noted at first in \cite{IK1}.
 According to Wiegmann \cite{W1,W2} the Kondo regime is realized in the
symmetric Anderson model in the integer valence state at $\rho v^2< -\epsilon_g$ (where $\rho$ is the density states of s-electrons at the Fermi energy).
We will consider the symmetric Anderson lattice defined on the chain  and simple square and cubic lattices.

\subsection*{Effective Hamiltonian of the symmetric Anderson model in the Kondo insulator state}

Using  $n^2_{\textbf{j},\sigma}=n_{\textbf{j},\sigma}$ we redefine the interaction term in the Hamiltonian (1) ${\cal H}_{int}= -\frac{1}{2}U\sum_\textbf{j}(n_{\textbf{j},\uparrow}-n_{\textbf{j},\downarrow})^2$.
The Hubbard-Stratonovich transformation converts the  problem with interaction into a non-interacting one in a stochastic field (hereinafter we will define
 as the $ \lambda $-field). We define the interaction term, taking into account the action ${S}_{0}$, defining the noninteracting terms in the Hamiltonian
   \begin{eqnarray}
   && S=S_{0}+2 \sum_{\textbf{j}} \frac{\lambda^2_{\textbf{j}}}{U}+2\sum_{\textbf{j}}\lambda_{\textbf{j}}(n_{\textbf{j},\uparrow}- n_{\textbf{j},\downarrow}).
    \label{A1}
   \end{eqnarray}
    The canonical functional is defined  as $${\cal Z}=\int {\cal D}[\lambda] \int {\cal D}[c^\dagger,d^\dagger,c,d] e^{-\beta S},$$ where the action $S=\frac{2}{U}\sum_{\textbf{j}}\lambda^2_{\textbf{j}}+
\int_0^\beta d\tau\Psi^\dagger (\tau)[\partial_\tau  + {\cal H}_{eff}]\Psi(\tau)$,  $\Psi (\tau)$ is  the wave function, $\beta=
\frac{1}{T}$, here $T$ is the temperature. Taking into account (2) we introduce an effective Hamiltonian ${\cal H}_{eff}={\cal H}_0 + 2\sum_{\textbf{j}}\lambda_{\textbf{j}}(n_{\textbf{j},\uparrow}-
   n_{\textbf{j},\downarrow})$  corresponding  to this action.
The translation invariance is conserved in an electron liquid state and as a result the value of  $ \lambda_{\textbf{j}} $ does not depend on $ \tau $.

The effective Hamiltonian includes the $Z_2$-field, associated with $\lambda_\textbf{j}$, where $\lambda_\textbf{j}$ could have two
   values $\pm \lambda$ at each lattice site. The energies of the d-electrons located at lattice site $\textbf{j}$ are equal to $-2\lambda_\textbf{j}$ and $2\lambda_\textbf{j}$.
   In this case, the local moment of the d-electron at each lattice site is not fixed, therefore s-electrons are hybridizes with d-electrons located randomly along the spin.
   According to Lieb\cite{Lieb}  a free configuration
   of a static $Z_2$-field corresponds to the energy minimum as a rule.
   A numerical calculations  of the energy of the Hamiltonian  ${\cal H}_{eff}$ showed  that the
   nontrivial uniform configuration of the $Z_2$-field corresponds to the energy minimum,
   a namely for $\lambda_\textbf{j}=\lambda$, $\lambda_{\textbf{j+1}}=-\lambda$.
  This uniform configuration forms the lattice with a double cell  at $\lambda\neq 0$ .

The  action $S$ is integrated out the fermion operators obtaining
\begin{eqnarray}
S(\lambda)=-T\sum_{\textbf{k}}\sum_n \sum_{\gamma=1}^{8} \ln [-i \omega_n+\varepsilon_\gamma(\textbf{k})]
+2N\frac{\lambda^2}{{U}},
 \label{A2}
\end{eqnarray}
where $\omega_n =T(2n+1)\pi$ are the Matsubara frequencies and
eight branches quasi-particle excitations $\varepsilon_\gamma(\textbf{k})$ ($\gamma =1,...,8$)
determine the electron states in an external magnetic field.

In the symmetric Anderson lattice describing by the effective Hamiltonian ${\cal H }_{eff}$ for $\epsilon_g=-\frac{U}{2}$,
the local density of d-electrons  constrain $n_{\textbf{k},\uparrow}+n_{\textbf{k},\downarrow}=1$ is valid in KIS
for an arbitrary value of the wave vector $\textbf{k}$  and  arbitrary parameters of the Hamiltonian (1), where
$n_{\textbf{k},\sigma}=\frac{1}{N}\sum_{\textbf{j}} n_{\textbf{j},\sigma}\exp (i \textbf{k}\textbf{j})$.
It is reduced to a local condition  $n_{\textbf{j},\uparrow}+n_{\textbf{j},\downarrow}=1$,  necessary for transformation of the local
spin-$\frac{1}{2}$ operator via the d-electron operators.

\section*{The ground-state}

In the  absence of an external magnetic field, the spectrum of quasi-particle excitations is doubly degenerated
in spin $\varepsilon_\gamma(\textbf{k})=\pm E_\pm(\textbf{k)}$ ($\gamma=1,...,4$)
\begin{equation}
 E^2_\pm(\textbf{k}) = \frac{1}{\sqrt2} [4 \lambda^2 +2 v^2+|w(\textbf{k})|^2 \pm \sqrt{16 \lambda^4 + 8\lambda^2 (2 v^2 - |w(\textbf{k})|^2) +
 |w(\textbf{k})|^2(4 v^2 + |w(\textbf{k})|^2)}],
 \label{eq:H5}
  \end{equation}
  where  $w(\textbf{k})=\sum^D(1+\exp(i k_\alpha))$, $\textbf{k}=(k_x,k_y,k_z)$ is the wave vector.

An external magnetic field shifts the branches of quasi-particle excitations and breaks the degeneracy of the spectrum, the spectrum splits into
eight branches of quasi-particle excitations  $\varepsilon_\gamma(\textbf{k})=\pm H \pm E_\pm(\textbf{k)}$ ($\gamma=1,...,8$).
At half filled occupation, the electron spectrum is symmetric with respect to zero energy and it is doubly degenerated in spin at $H=0$ (4).
The electron spectrum consists of  eight branches of quasi-particle excitations.
The chemical potential is equal to zero. For $v,\lambda \neq 0$ it is gapped, the gap opens at $\textbf{k}=0$ for $v>2\sqrt {2}\lambda$ or at $\textbf{k}=\overrightarrow{\pi}$ for $v <2\sqrt {2}\lambda$. The first case corresponds to intermediate valence regime of local d-states. We consider the case $v<2\sqrt {2}\lambda$, in which a local moment $m=\frac{1}{2N}\sum_{\textbf{k}}(n_{\textbf{k},\uparrow}-n_{\textbf{k},\downarrow})$ is changed in the interval $[\frac{1}{6},\frac{1}{2}]$, where $m =\frac{1}{2}$ for $v=0$ and $m=\frac{1}{6}$ for $v\to \infty$.

KIS arises as a result of flip up spin scattering of $s$-electrons on the local moments.
The condition for the realization of such phase state is considered in the symmetric Anderson lattice .
The local moments of $d$-electrons are realized for
integer valence state at $v<2\sqrt {2}\lambda$
for arbitrary values of $v$ and $\lambda$. The result of numerical calculation $m$ for a square lattice is shown in Figure 1.
 The behavior of $m$  has an universal form for different dimensions in the coordinates $(\lambda,v)$
 ( the surface in  Figure 1 is slightly deformed in different dimensions).
 The small value of $m$ corresponds to the absence of local moments in the intermediate valence of d-states.
The value of the gap is an universal function of $v, \lambda $  for arbitrary dimension \cite{IK1}  and  is equal to
$\Delta (H)= \Delta (0)-2H, \Delta(0)=2(-\lambda  +\sqrt{\lambda^2 +  v^2})$, in the weak $v$-limit  $\Delta (0) \simeq \frac{v^2}{\lambda}$.

In  the magnetic field $H<H_c=\frac{1}{2}\Delta(0)$ the phase state of the electron liquid
with $\sum_j <c^\dagger_{j,\uparrow} c_{j,\uparrow}>=\sum_j<c^\dagger_{j,\downarrow} c_{j,\downarrow}>$ is gapped and
the ground state energy does not depend on the magnetic field.
In saddle point approximation a magnitude of the $\lambda$-field does not depend on the  magnetic field, in the ground state
the self-consistent equation has the following form
\begin{equation}
\frac{4\lambda}{U}= \frac{1}{2N}\sum_{\textbf{k}}\int d \textbf{k} \left(\frac { \partial E_+(\textbf{k})}{\partial \lambda}+
\frac{ \partial E_-(\textbf{k})}{\partial \lambda} \right).
\label{eq:H6}
\end{equation}

The self-consistent equation makes it possible to determine the magnitude of the $\lambda$-field depending on the parameters of the model Hamiltonian (1)
$v$ and $U$.
The calculations of $\Delta (0)$ are shown in the coordinates $U,v$ for different dimensions of the model in Figures 2. According to the numerical calculations the
value of $\Delta (0)$ changes slightly whith  the model dimension changing.

An external magnetic field breaks the spectrum  degeneracy causing the spectrum splitting into eight branches of quasi-particle excitations.
The value of gap decreases with increasing magnetic field and the gap closes at a critical value $H_c $.
The spectra  of the quasi-particle excitations of  an electron liquid calculated for
the chain and square lattice correspondingly (where at $H=0$ $\Delta=0.1656$ and at $H_c=0.0828$ $\Delta=0$)  are shown in Figs 3 and Figs 4.
The effect of  magnetic field on the phase state of the electron liquid is reduced to a decrease in the gap.
For $H>H_c$ the quasi-particle spectrum is
gapless and KIS disappears. The behavior of electron liquid in the small magnetic fields $H<H_c$
at which KIS realized is studied.
The nature of the Kondo insulator and topological KISs is similar and
the topological states define the surface properties of compounds while the bulk their properties are  the same usually.

\section*{Conclusion}

The symmetric Anderson lattice at half-filling for different dimensions is studied.
We propose a new physical mechanism for the formation of KIS using the mean field approach .
The behavior of electron liquid in KIS is similar to  the Majorana spin liquid behavior in the Kitaev model.
The local moments of d-electrons form a static $Z_2$-field in symmetric Anderson lattice where band electrons move. At half filled occupation
the ground state configuration of this field corresponds to the lattice with a double cell. In contrast to  the gapless Majorana spin liquid,
an electron liquid in KIS is gapped.
The scattering of band electrons on $d$-electrons depends on their spins In the Kondo problem and the spin flip scattering dominates and forms
the Abrikosov-Suhl resonance.
Due to a static $Z_2$-field $s$-electrons hybridize with $d$-electrons with energies  determined by the  field.
This one-particle consideration makes it possible  to take into account the interaction between $s-$ and $d-$electrons in the framework
the Hamiltonian that does not break the symmetry of the model for arbitrary dimension.

In one-particle effective Hamiltonian s- and d-electrons  are hybridized with the same spins,
in this case, a static $Z_2$-field changes the energies of the local d-states with which the band electrons are hybridized.
Thus, the band electrons  are hybridized successively with different local d-states.
In the symmetric Anderson model at half filling, the spectrum of quasi-particle excitations is symmetric with respect to zero energy,
it is similar to Majorana's spectrum.
The value of the gap is determined by both $v$ and $U$ and the gap protects KIS.
 There is a critical value of the magnetic field at which the gap closes and this way a phase state is destroyed.

\section*{Acknowledgments}
The work was supported by the program competition theme "Physical development of materials for elastic-caloric cooling (ELAST)" No 0122U001169 .
The author thanks to Springer Nature waivers team for support of open access publication of the paper.

\section*{Author contributions statement}
I.K. carried out analytical and numerical calculations, prepared the manuscript, figures

\section*{Additional information}
The author declares no competing financial interests. \\
\section*{Availability of Data and Materials}
All data generated or analysed during this study are included in this published article.\\
Correspondence and requests for materials should be addressed to I.N.K.

\end{document}